\documentclass[aps,twocolumn]{revtex4}
\def\BibTeX{{\rm B\kern-.05em{\sc i\kern-.025em b}\kern-.T
    08em\kern-.1667em\lower.7ex\hbox{E}\kern-.125emX}}
\usepackage{graphicx}

\begin{document}
\title{Exact cloaking of uniform magnetic fields with a superconductor-magnetic bilayer}

\author{Carles Navau, Jordi Prat-Camps, and Alvaro Sanchez}
\affiliation{Grup d'Electromagnetisme, Departament de F\'{\i}sica, Universitat Aut\`onoma de Barcelona, 08193 Bellaterra,
Barcelona, Catalonia, Spain
}

\begin{abstract}

The experimental realization of cloaks for electromagnetic fields has been possible only in reduced schemes because of  extreme material conditions required in exact cloaks. Here we analytically demonstrate that a simple arrangement of two concentric homogeneous and isotropic layers with different permeabilities can keep a uniform externally applied magnetic field unperturbed. If the inner layer is superconducting, the device prevents exterior fields from entering an enclosed region, which represents a case for an exact cloaking of uniform fields. When the applied field is inhomogeneous the cloaking property is reduced but may be optimized by changing the thickness of the layers.

\end{abstract}

\maketitle

The possibility of cloaking electromagnetic waves has opened a whole field in physics after being theoretically predicted in 2006 \cite{controlling,leonhardt}, with some precursor works appeared earlier \cite{greenleaf}. It was soon realized that the proposed cloaking schemes involved very extreme parameters, such as permittivities $\varepsilon$ or permeabilities $\mu$ that are anisotropic and at the same time very inhomogeneous (heavily dependent on position) and/or with diverging (singular) values  \cite{chen_natmat}. Because of this, in order to implement these cloaking ideas into practice some reductions and simplifications have been needed \cite{cai,collins,yan}, giving up exact cloaking. Some examples are the reduced scheme used for microwaves in \cite{schurig} and the carpet cloak approach \cite{carpet}, the latter being the basis for some implementations in the optical range \cite{chen,3D}. Therefore, all experimental realizations are based not on exact cloaking proposals but on partial cloaking schemes; exact designs involve material properties that are far from being feasible \cite{chen_natmat,leonmaxwell}.

This work addresses a main question: Can a exact (non-reduced) simple cloak being designed, one made only of homogeneous isotropic non-singular media? It seems at first that the answer is negative, since cloaking designs are based on space transformations, which result in anisotropy and/or singular values of the involved $\varepsilon$ or $\mu$ values. For example, the original cylindrical cloak introduced in \cite{controlling} results from a transformation of space consisting in blowing out a line into a cylindrical volume, which yields anisotropic tensors $\varepsilon$ or $\mu$ that have singularities at the inner shell of the cylindrical volume\cite{controlling,chen_natmat}.

There is a particular case of cloaking that deserves a separate consideration, that for static magnetic fields, involving the concept of dc metamaterials \cite{wood,ourAPL,magnus,antimagnet}. Actually, controlling magnetic fields is a crucial aspect of physics and engineering. The particular case of achieving magnetic cloaking would be indeed a most interesting result. Because no distortion of external field is equivalent to non-detectability, if this possibility is realized, one could place magnetic objects inside a cloak and render them magnetically undetectable from the exterior. However, it was found by Wood and Pendry \cite{wood} that the required permeability values are similar to those required for waves -with the main difference that for static fields electric and magnetic effects decouple and one needs to care only of the permeabilities. Therefore, perfect cloaking with non-singular and isotropic parameters seems apparently impossible in the static case as well.

Is it then true that designing an exact (not reduced) cloak with homogeneous, isotropic and non-singular parameters is impossible, even for static magnetic fields? In this work, we will show that such a possibility exist, by analytically demonstrating that a simple structure of two concentric homogeneous isotropic layers with selected values of magnetic permeabilities (one diamagnetic and another paramagnetic) can result in an exact null distortion of uniformly applied magnetic fields. When the inner layer is a superconducting one, then not only the external fields are non-distorted but a null magnetic field in the enclosed central region is also exactly achieved; these are precisely the two conditions for a perfect cloaking behavior. 
We will also show numerically that even when the field is not spatially uniform, a reasonable good cloaking behavior can be obtained, and will indicate some ways to improve this approximate cloaking in real situations.

How such an exact cloaking with the simplest parameters can be obtained in the static magnetic case? The key physical idea lies on the magnetic response of the components of the cloaking. A superconducting shell would completely shield an enclosed volume from a uniform external magnetic field, at the price of distorting that field in its exterior; roughly speaking, it 'expels' {\bf B} lines. A (para)magnetic material has exactly the opposite response, 'attracting' {\bf B} lines; the larger its permeability the more concentration of {\bf B} lines it provides. Both cases are illustrated in Fig. 1(c) and (d). We shall prove in this work that both effects can be exactly canceled for a uniform applied field when considering an adequate superconducting-magnetic bilayer. Actually, a whole family of solutions exists which relates the permeability of the magnetic material with the dimensions of the bilayer, as we show next.

Consider an infinitely long (along $z$-direction) cylindrical magnetic shell of interior and exterior radii $R_0$ and $R_1$, respectively. This shell is surrounded by another cylindrical shell of radii $R_1$ (interior) and $R_2$ (exterior). We assume that both magnetic materials are homogeneous, isotropic, linear, characterized by uniform relative permeabilities $\mu_1$ for the inner shell and $\mu_2$ for the outer one. We call this system a bilayer.
Consider a uniform externally applied magnetic field directed along the $x$-direction, ${\bf H}_a=H_a \hat{x}$. The magnetic field distribution in the whole space can be analytically calculated, as shown in the Supplementary Material. The scalar potential in the outer region ($\rho>R_2$) is
\begin{eqnarray}
	\label{eq.phi3}\phi_3&=&\left(-H_a \rho + {A\over \rho} \right) \cos\theta,
\end{eqnarray}
where $A$ is a constant determined by boundary conditions and the parameters 
$\mu_1, \mu_2, R_0, R_1,$ and $R_2$.
The term with $A/\rho$ in $\phi_3$ [Eq. (\ref{eq.phi3})] can be viewed as the modification of the applied magnetic potential due to the bilayer. Actually, we could calculate the magnetic field created by a single shell as a particular solution of the above results. For example, in fig. 1(d) we show the case of a superconducting shell ($\mu_1=0$ and $\mu_2=1$) and in fig. 1(c) a single (para)magnetic shell ($\mu_1=\infty$ and $\mu_2=1$).

We now arrive to the key point. Since the distortion of both layers has the same radial dependence outside the bilayer, both contributions have merged in just one constant, the $A$ in Eq. \ref{eq.phi3}. Thus, by setting $A=0$ we find the combination of $\mu_1$ and $\mu_2$ that cancels the distortion of the uniform applied field.
This non-distortion condition yields a general combination of positive permeabilities for the cylindrical case [given in the Supplementary Material, Eq. (22)].

Of special interest is the particular case in which the inner layer is an ideal superconductor, with $\mu_1=0$. In this case, the non-distortion condition for the bilayer reduces simply to 
\begin{equation}
  \label{eq:mu2WithSC}
	\mu_2={R_2^2+R_1^2 \over R_2^2-R_1^2}.
\end{equation}
The ${\bf B}$-field lines for a bilayer with an inner ideal superconducting layer, $\mu_1=0$, and an outer layer with a permeability given by Eq.(\ref{eq:mu2WithSC}) are shown in Figs. 1(a) and (b), where we observe how the field lines are excluded from the interior of the ideal superconductor, whereas they are concentrated in the outer magnetic layer. Since the combination of permeabilities satisfies the non-distortion condition, the field lines at the exterior region are exactly those corresponding to only the applied field.

It is also interesting to note that any pair of permeabilities satisfying the non-distortion condition [Eq. (22) in the Supplementary Material]
will produce a uniform field inside the hole. The strength of the magnetic field in that region is zero only for the cases $\mu_1=0$ or $\mu_2=0$ or $\mu_1\rightarrow \infty$ or $\mu_2\rightarrow \infty$, i. e., wherever at least one of the layers is either an ideal superconductor or an ideal soft ferromagnet.

Similar analytical derivations can be made for a spheric configuration, instead of a cylindrical one. A hollow sphere with a hole of radius $R_0$, an inner layer with radius $R_0$ and $R_1$ and permeability $\mu_1$, and an outer layer with radius $R_1$ and $R_2$ with permeability $\mu_2$ ($R_0<R_1<R_2$) produces no external field distortion by a relation given in the Supplementary Material [Eq. (39)]. Magnetic fields in all regions can be obtained analytically also in this case. 
When the inner layer is an ideal superconductor, the non-distortion in the exterior regions requires
\begin{equation}
  \label{eq:mu2WithSCSph}
	\mu_2={2R_2^3+R_1^3 \over 2(R_2^3-R_1^3)}.
\end{equation}

We have proven so far that cylindrical and spherical superconductor-magnetic bilayers can exactly cloak a uniform magnetic field, if the adequate parameters are chosen. In most applications, however, it would be interesting to have some cloaking behavior (even an approximate one) when the applied field has some spatial variation. We next provide a way to estimate the distortion in the non-uniform field case and a general procedure to reduce this distortion.

The magnetic response of a superconducting-magnetic bilayer to the fields of a single small magnet (basically a dipole field) at different distances from the bilayer are shown in Fig. 2, calculated by Comsol Multiphysics software, using the electromagnetics module (magnetostatics). In these cases of non-uniform applied field, a distortion of the external field occurs - distortion gets larger with increasing applied field inhomogeneity-, but it can be seen that this distortion is less than few percent in all space except in a region close to the field source and the bilayer \cite{footnote}.
As an example, the dark pink region corresponding to a 5\% distortion that in the case of a dipole at a distance 2.5$R_2$ occupies a region of about 2-3$R_2$, which in the case of a single superconducting layer or a single magnetic one [like those in Figs. 1(c) and 1(d)] extends to more than 15$R_2$.

There is an important effect worth to discuss. When \textit{reducing} the thickness of the outer magnetic layer  $R_2$, keeping $R_1$ constant, the region in which the external field is distorted is reduced. Actually, for thin magnetic layers, $R_2-R_1<<R_1$, Eq. \ref{eq:mu2WithSC} reduces to $\mu_2\simeq R_1/(R_2-R_1)$. In that case, the large values of $\mu_2$ allow a strong modification of the external field inside the thin layer. This effect is seen in Fig. 3, where the calculated distortion zone in the case of a near dipole, an unfavorable case in terms of field inhomogeneity, is clearly reduced by decreasing the thickness of the magnetic layer.

It it interesting to notice that in order to obtain a real cloak behavior and not only a non-distorted external field, then the order of the layers is relevant, and the superconducting layer has to be the inner one. It is this layer that actually decouples magnetically the interior from the exterior, the role of the outer magnetic layer being simply to eliminate (or reduce for inhomogeneous applied fields) the distorting effect of the inner one. This remarkable property of superconductors in the magnetostatic case ($\mu=0$) has no simple equivalent for the general electromagnetic case. For this reason, it is likely that a system with only homogeneous, isotropic and non-singular values of $\mu$ and $\varepsilon$ cannot be found for exactly cloaking  electromagnetic waves.

Finally, our results may have not only conceptual but practical implications as well, in particular in fields like reducing the magnetic signature of vessels or in allowing patients with pacemakers or cochlear implants to be allowed to use medical equipment based on magnetic fields, such as magnetic resonance imaging MRI \cite{mri} or transcraneal magnetic stimulation \cite{trans}. For practical implementations, one could use a superconducting material whose critical field is less than the range of fields involved (even type-II superconductors with high critical current would have an adequate behavior \cite{araujo}), and, in order to have a homogeneous material with positive $\mu$, one could use, for example, a comercial ferromagnetic material for transformer core, as in \cite{gomory}. Provided that the materials are in the linear range, our design is passive and lossless; for this reason, in practice, the proposed implementation would work mainly at low applied magnetic fields. Moreover, ways to discretize the magnetic layers in a cloaking device for magnetic fields, such those discussed in \cite{antimagnet} can be used here.

We acknowledge D.-X. Chen for valuable comments and Consolider Project NANOSELECT (CSD2007-00041) for financial support.

\newpage

\begin{figure}
	\centering
		\includegraphics[width=0.4\textwidth]{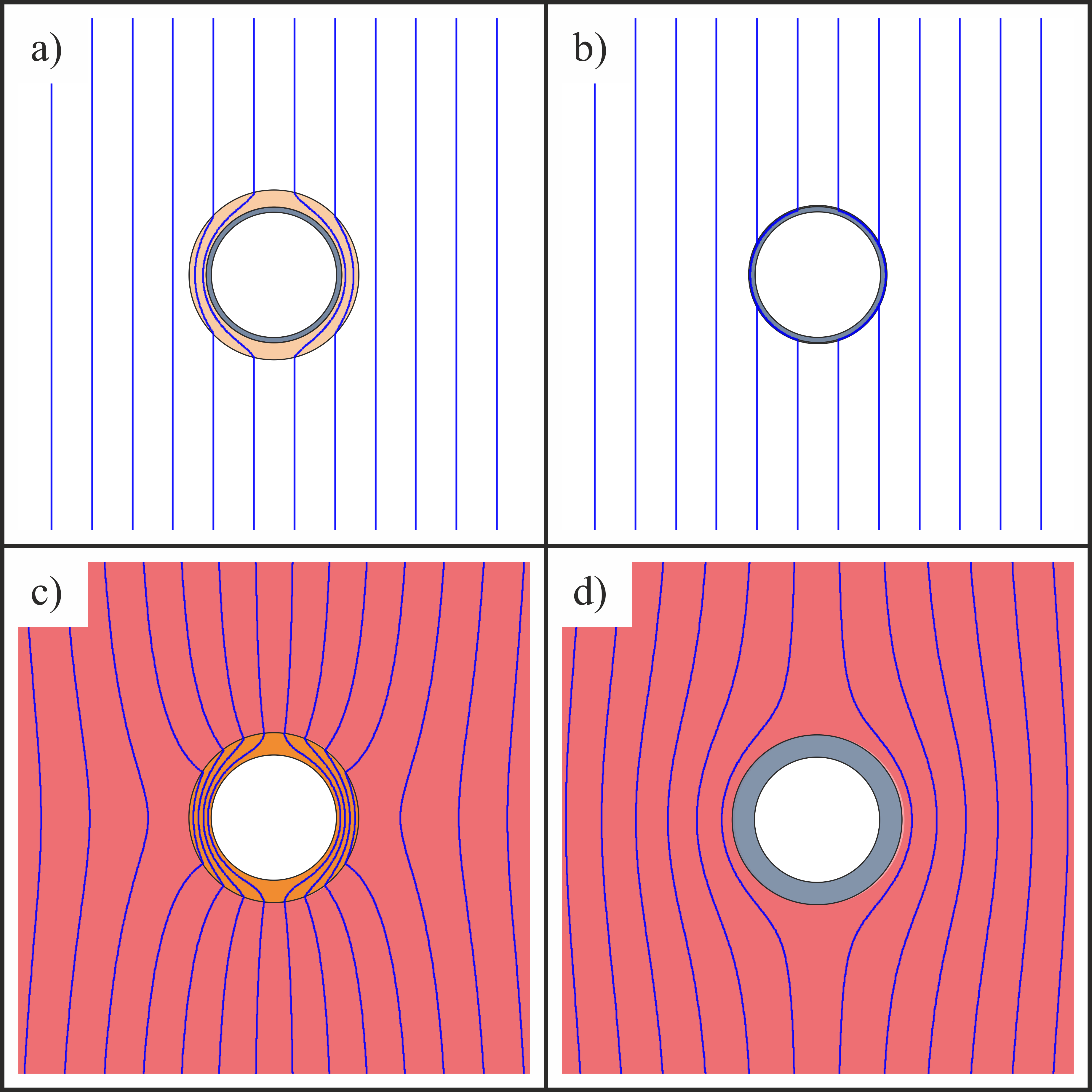}
\caption{Calculated field lines for (a) a cylindrical bilayer with an inner superconducting layer ($\mu=0$) of interior (exterior) radius of  $R_0=0.95R_1$ ($R_1$) and an outer magnetic layer with $R_2/R_1=1.25$ with $\mu_2=4.556$, (b) a similar bilayer with $R_2/R_1=1.025$ and $\mu_2=40.506$. For comparison we show in (c) and (d) a single cylindrical shell with $\mu=\infty$ and $\mu=0$, respectively, using the same thickness as the bilayer in (a).}
\end{figure}

\begin{figure}
	\centering
		\includegraphics[width=0.4\textwidth]{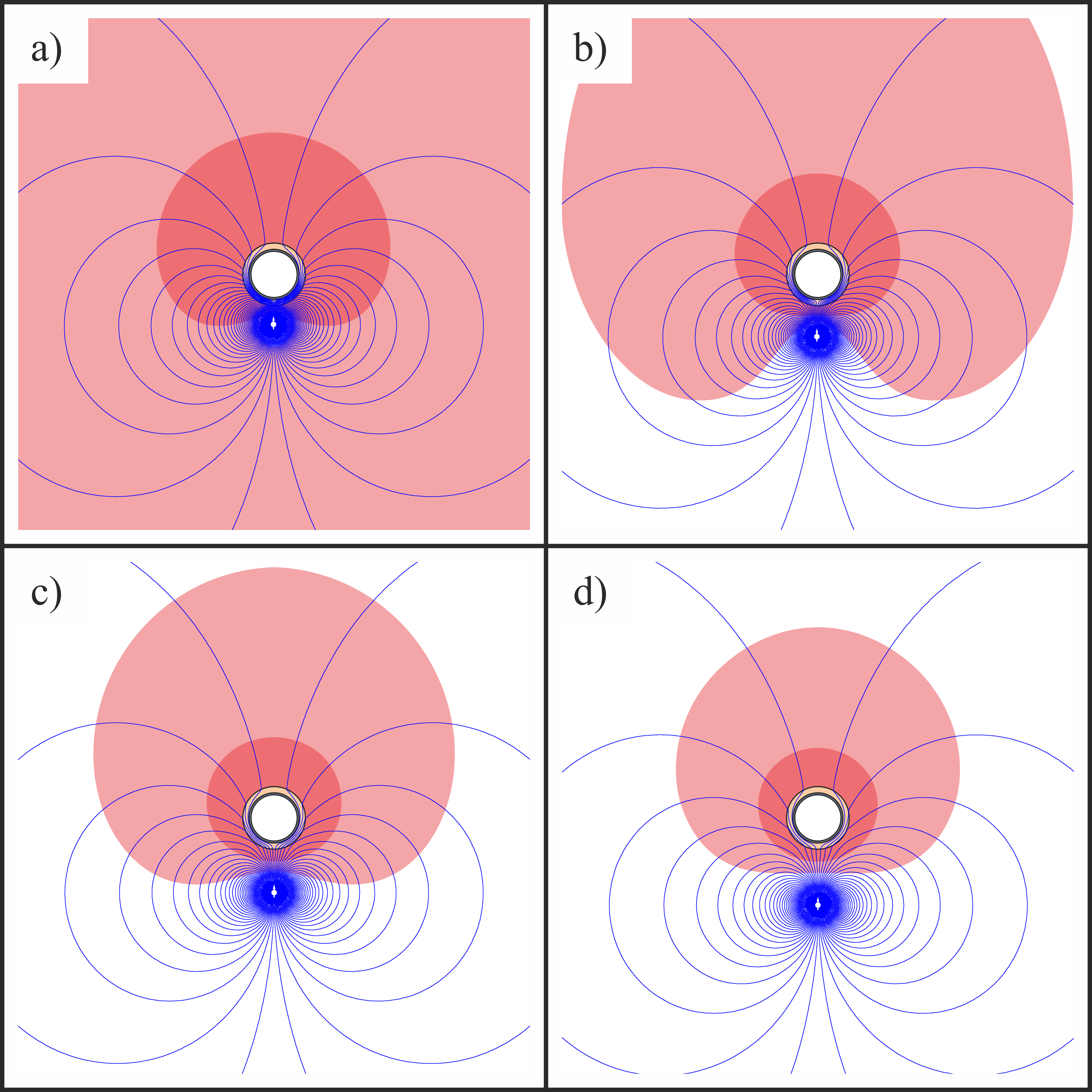}
\caption{Calculated field lines for the interaction of a bilayer device (with $R_2/R_1=1.25$) with the field created by a magnetic dipole placed at 
different distances from the center of the device: (a) $2R_1$, (b) $2.5R_1$, (c) $3R_1$ and (d) $3.5R_1$. Bright (dark) pink-shaded regions indicate where distortion is larger that $1\%$ ($5\%$). }
\end{figure}

\begin{figure}
	\centering
		\includegraphics[width=0.4\textwidth]{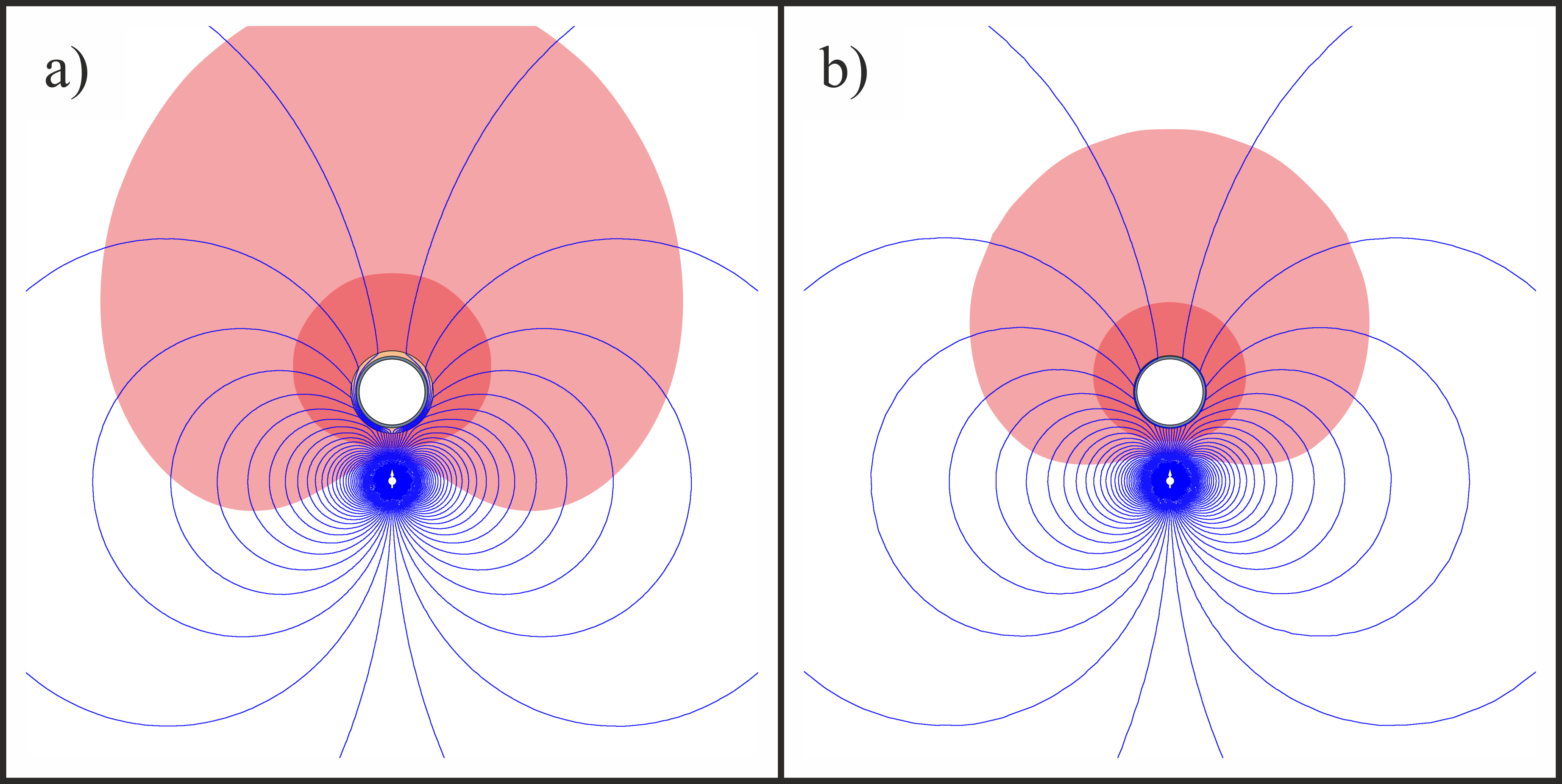}
\caption{Calculated field lines and distortion regions for the interaction of a dipolar field (source is at a distance of $2.5R_1$ from the center of the bilayer) with bilayer devices with different thickness for the outer layer [(a) $R_2/R_1=1.15$, and (b) $R_2/R_1=1.025$], with corresponding permeabilities given by Eq. \ref{eq:mu2WithSC}. Bright (dark) pink-shaded regions indicate where distortion is larger than 1\% (5\%).}
\end{figure}


\begin{thebibliography}{99}


\bibitem{controlling} J. B. Pendry, D. Schurig, and D. R. Smith, Science {\bf 312}, 1780 (2006). 

\bibitem{leonhardt} U. Leonhardt, Science {\bf 312}, 1777 (2006). 


\bibitem{greenleaf} A. Greenleaf, M. Lassas, and G. Uhlmann, Physiol. Meas. {\bf 24}, 413 (2003). 


\bibitem{chen_natmat} H. Chen, C. T. Chan, and P. Sheng, Nat. Mat. {\bf 9}, 387 (2010).


\bibitem{cai} W. Cai, U. K. Chettiar, A. V. Kildishev, and V. M. Shalaev, Nat. Photonics {\bf 1}, 224 (2007).

\bibitem{collins} P. Collins and J. McGuirk, J. Opt. A: Pure Appl. Opt. {\bf 11}, 015104 (2009).

\bibitem{yan} M. Yan, Z. Ruan, and M. Qiu, Optics Express {\bf 15}, 17772 (2007).


\bibitem{schurig} D. Schurig, J. J. Mock, B. J. Justice, S. A. Cummer, J. B. Pendry, A. F. Starr, and
    D. R. Smith, Science {\bf 314}, 977 (2006).


\bibitem{carpet} J. Li and J. B. Pendry, Phys. Rev. Lett. {\bf 101}, 203901 (2008).

\bibitem{chen} X. Chen, Y. Luo, J. Zhang, K. Jiang, J. B. Pendry, and S. Zhang, Nat. Commun. {\bf 2}, 176 (2011).



\bibitem{3D} T. Ergin, N. Stenger, P. Brenner, J. B. Pendry, and M. Wegener, Science {\bf 328}, 337 (2010).



\bibitem{leonmaxwell} U. Leonhardt, Science {\bf 471}, 293 (2011). 



\bibitem {wood} B. Wood and J. B. Pendry, J. Phys. Condens. Matter {\bf 19}, 076208 (2007). 

\bibitem{ourAPL} C. Navau, D.-X. Chen, A. Sanchez, and N. Del-Valle, Appl. Phys. Lett. {\bf 94}, 242501 (2009).

\bibitem{magnus} F. Magnus, B. Wood, J. Moore, K. Morrison, G. Perkins, J. Fyson, M. C. K. Wiltshire, D. Caplin, L. F. Cohen, and J. B. Pendry, Nature Materials {\bf 7}, 295 (2008).

\bibitem{antimagnet} 
A. Sanchez, C. Navau, J. Prat-Camps, and D.-X. Chen, New J. Phys. {\bf 13}, 093034 (2011).


\bibitem{jackson}
J. D. Jackson, {\sl Classical Electrodynamics} (3rd ed., Wiley, New York), 1999.

\bibitem{footnote} The distortion zone showing how a magnetic induction {\bf B} differs from the externally applied magnetic induction ${\bf B}_{\rm ext}$ in less than $p$ per cent is calculated as the points in space following the condition
$
{\mid {\bf B}-{\bf B}_{\rm ext}\mid /\mid {\bf B}_{\rm ext}\mid}>p/100
$.

\bibitem{mri} A. Roguin, J. Am. Coll. Cardiol. {\bf 54}, 556 (2009).


\bibitem{trans} M. Kobayashi and A. Pascual-Leone, Lancet Neurology {\bf 2}, 145 (2003).

\bibitem{araujo} F. M. Araujo-Moreira, C. Navau, and A. Sanchez, Phys. Rev. B {\bf 61}, 634 (2000).

\bibitem{gomory} M. Vojenciak, J. Souc, and F. Gomory, Supercond. Sci. Technol. {\bf 24},
075001 (2011).



\end{thebibliography}
\end{document}